\documentclass[]{revtex4-1}
\usepackage[utf8]{inputenc}
\usepackage{amsmath}
\usepackage{natbib}
\usepackage{graphicx}
\usepackage{caption}
\usepackage{subcaption}

\newcommand{\beq}{\begin{equation}}
\newcommand{\eeq}{\end {equation}}
\newcommand{\bea}{\begin{eqnarray}}
\newcommand{\eea}{\end{eqnarray}}

\begin{document}


\title{Chirality-enabled optical dipole potential energy for two-level atoms}

\author{V. E. Lembessis}
\affiliation{%
Quantum Technology Group, Department of Physics and Astronomy,
College of Science, King Saud University, Riyadh 11451, Saudi Arabia}


\author{K. Koksal}
\affiliation{
Physics Department, Bitlis Eren University, Bitlis 13000; Turkey}
\author{J. Yuan and M. Babiker}
\affiliation{
Department of Physics, University of York, YO10 5DD, UK}


\date{\today}

\begin{abstract}
We consider the optical dipole potential energy, which arises from the interaction of a two-level atom with a circularly polarized Laguerre-Gaussian laser beam of small waist. The beam is characterized by the existence of a longitudinal electric field component which is responsible for the appearance of a chiral term in the optical dipole potential energy. This term reverses sign if either the winding number or the wave polarization of the beam reverses sign. We propose a scheme of a bi-chromatic vortex interaction with the two-level atom in which the resulting optical dipole potential is fully chiral.  
\end{abstract}

\maketitle

Optical vortices comprise a large family of laser beams in which each member carries an orbital angular momentum (OAM) along the propagation direction \cite{Andrews2012}. This OAM is quantized in units of $\ell\hbar$ where $\ell$ is a non-zero integer number. The photons of a circularly polarized vortex beam may also have a spin angular momentum (SAM) which has two possible states, corresponding to right- and left-hand polarizations. The most prominent member of the family of optical vortices is the Laguerre-Gaussian (LG) beam \cite{Allen1992}. 

 When an LG beam is created with a very small beam waist $w_0$, a longitudinal (or axial) electric field component appears. The relative magnitude of this component is small for relatively large $w_0$ and this is the reason why its effects have so far been ignored in various theoretical and experimental works especially those dealing with the mechanical effects of laser radiation pressure on atoms \cite{Babiker2019}.  But a small beam waist, on the one hand, leads to a considerable increase in the size of this axial field component, which becomes comparable to the transverse components. The significance of this longitudinal electric field component has not only to do with its relevant size - it brings into consideration the concept of chirality in light-matter interactions for beams of such a small waist. In this regime optical spin couples to the spatial degrees of freedom of the light through a spin-orbit interaction (SOI) and gives rise to significant effects. Over the last few years there has been a rapidly growing interest in the community in these effects. We are familiar with SOI effects in relativistic mechanics but not so much with the analogous effects that exist in the physics of light beams. However, the SOI arises normally from the solution of Maxwell’s equations under tight focusing conditions and they become important at the sub-wavelength scales in plasmonics and nanophotonics \cite{Bliokh2015}. Note that beams with small waists have enhanced beam convergence effects embodied in the Gouy and curvature phase in the vicinity of the focal plane.   

 The SOI interaction plays an important role in the mechanical effects of light on atoms and on tiny particles and, in particular, in radiation pressure forces exerted on them by beams with very small beam waists. In 1996 Allen et al showed for the first time that when a two-level atom interacts in free-space with a circularly polarized LG beam, the azimuthal component of the scattering force depends on the coupling between the spin of the beam and its OAM.  It was also shown (for the first time) that the polarization of light can also affect the gross motion of the atom and not simply its internal dynamics \cite{Allen1996}. In their paper the authors showed that the SOI contribution is typically very small but they did not realise that it could be of considerable size in beams with very small beam waists. Since then investigations of the SOI effects have not focused on its role in the mechanical effects on atoms i.e. on the modification of the translational atomic motion but there have been works on its mechanical effects on small particles \cite{Cao2016}. In a recent report by Quinteiro et al the role of the longitudinal electric field component was investigated in the quadrupole interaction of a calcium ion situated in the dark core of an LG beam  \cite{Quinteiro2017}. In another report the existence of this field component was probed with the help of the interaction of a single molecule with a fixed dipole moment \cite{Novotny2001}. In a very recent paper the creation of a chiral optical dipole potential energy is shown in a scheme which involves the irradiation of chiral molecules with three femto-second laser beams which ensures a cyclic three-level configuration\cite{Kazemi2020}. However, in \cite{Quinteiro2017} the chiral potential energy arises from the chiral nature of the transition matrix elements. In our work here the transition matrix element is polarization-independent and also involves a simple two-level scheme.

In this Letter we report the findings of the calculation of the optical dipole potential energy experienced by a two-level atom when it interacts far off-resonantly with a circularly polarized LG beam of very small beam waist and we  also point out the existence of a strongly chiral term in the optical dipole potential. The structure of the paper is as follows. We first present the description of the atom-beam interaction and derive the exact formula for the corresponding Rabi frequency which leads to the desired expression for the optical dipole potential energy.  We then present a case where a purely chiral optical dipole potential energy can be realized when a two-level atom interacts with a bi-chromatic vortex beams' arrangement. We demonstrate the existence and the relevant size of this chiral potential energy with numerical examples based on experimentally accessible parameters. 
	
 The electric field of a circularly-polarized LG$_{p,\ell}$ beam of small beam waist can be written as follows \cite{Allen1999}:
\begin{widetext}
\begin{equation}
    \vec{E}=\frac{1}{2}\left[\alpha E\vec{x}+\beta E\vec{y}+\frac{i}{k}\left(\alpha\frac{\partial E}{\partial x}+\beta\frac{\partial E}{\partial y}\right)\vec{z}\right]e^{(ikz-i\omega t)}
\end{equation}
where $\alpha$ and $\beta$ are in general complex constants.  It is convenient to write the function $E$  in cylindrical polar coordinates ${\bf r}=(\rho,\phi,z)$ as follows
\begin{equation}
    E=u\exp{(i\Theta)}
\end{equation}
\begin{equation}
    u=\frac{E_0C_{p,|l|}}{\sqrt{1+z^2/z_R^2}}\left(\frac{\rho \sqrt{2}}{w_0\sqrt{1+z^2/z_R^2}}\right)^{|\ell|}\exp{\left[-\frac{2\rho^2}{w_0^2(1+z^2/z_R^2)}\right]}L_p^{|\ell|}\left(\frac{2\rho^2}{w_0^2(1+z^2/z_R^2)}\right)
\end{equation}
\begin{equation}
    \Theta=\ell\phi-(2p+|\ell|+1)\arctan{\left(\frac{z}{z_R}\right)} +\frac{kz\rho^2}{2(z^2+z_R^2)}
\end{equation}
With the above notation the vortex electric field can be written as 
\begin{equation}
      \vec{E}=\frac{1}{2}\left\{\alpha E\vec{x}+\beta E\vec{y}+\frac{1}{k}\left[i\left(\alpha\frac{\partial u}{\partial x}+\beta\frac{\partial u}{\partial y}\right)-u\left(\alpha\frac{\partial \Theta}{\partial x}+\beta\frac{\partial \Theta}{\partial y}\right)\right]\vec{z}\right\}e^{ikz+i\Theta-i\omega t}
\end{equation}
In the above $E_0$ and $C_{p,|\ell|}$ are constants, $k$ is the axial wavevector while $w_0$ is the beam waist at the focal plane $z=0$ and $z_R$ is the Rayleigh range.  Note that the overall phase function includes the Gouy and curvature phases.
The intensity of the beam is associated with the modulus squared of this field given by:
\begin{align}
    |\vec{E}|^2 =\vec{E}*\vec{E}^*=&\frac{1}{4}\left(\alpha^2+\beta^2\right)u^2-\frac{i\sigma \ell u}{4k^2\rho}\frac{\partial \Omega}{\partial \rho} \nonumber \\
    &+\frac{1}{4k}\left[|\alpha|^2\cos^2{\phi}\left(\frac{\partial u}{\partial \rho}\right)^2+|\beta|^2\sin^2{\phi}\left(\frac{\partial u}{\partial \rho}\right)^2+|\alpha|^2u^2\left(\frac{\ell\sin{\phi}}{\rho}\right)^2+|\beta|^2u^2\left(\frac{\ell\cos{\phi}}{\rho}\right)^2\right]
\label{eq:beam_intensity}.
\end{align}
\end{widetext}
where for a circularly polarized beam we have $\sigma=\alpha \beta^*-\alpha^* \beta$ and $\alpha \beta^*+\alpha^* \beta=0$. Clearly, as Eq. \ref{eq:beam_intensity} shows, the intensity of the beam includes a term, which couples the spin of the beam to its OAM. The sign of this term can be changed either by varying the polarization sense or by changing the OAM winding sense. This chiral term is carried over in the interaction of the beam with a two-level atom. Let us assume the case where we work close to the beam focal plane situated at $z=0$ and the interaction of the two-level atom with the above electric field is at far-off resonance. In this case due to the very large detuning the atomic fine-structure is not resolved  
\cite{Grimm2000}. We can then associate the beam intensity with a Rabi frequency, the square of which is given by:
\begin{equation}
    \tilde{\Omega}=\sqrt{\Omega^2+\left[\frac{1}{k\sqrt{2}}\left(\frac{\ell\Omega}{\rho}-i\sigma \frac{\partial \Omega}{\partial \rho} \right)\right]^2}
    \label{eq:beam_intensity_Rabi_frequency}.
\end{equation}
where $\Omega=\Omega_0C_{p,|\ell|}\left(\frac{\sqrt{2}\rho}{w_0}\right)^{|\ell|} \exp{\left(-\frac{\rho^2}{w_0^2}\right)}L_p^{|\ell|}\left(\frac{2\rho^2}{w_0^2}\right)$ and we have taken into account the fact that in a circularly polarized beam $|\alpha|=|\beta|=1/\sqrt{2}$ and $\sigma=\pm i$. In other words, the beam-atom interaction has an effective Rabi frequency which is the contribution of a Rabi frequency due to an ordinary circularly polarized LG beam and a Rabi frequency which corresponds to the longitudinal electric field term. As can also be seen from Eq.\ref{eq:beam_intensity} 
the contribution of the spin-orbit coupling in the Rabi frequency is $\frac{i\sigma \ell\Omega}{k^2\rho}\frac{\partial \Omega}{\partial \rho}$. This is directly proportional to the term $\frac{\partial \Omega}{\partial \rho}$. But this term is given by $\frac{\partial \Omega}{\partial \rho}=\Omega\left(\frac{|\ell|}{\rho}-\frac{2\rho}{w_0^2}\right)$ (when $p = 0$) and becomes zero when $\rho=w_0\sqrt{\ell|/2}$, i.e. at points where the intensity of an ordinary (i.e without the contribution of the electric field component along $z$) LG beam maximizes \cite{Babiker2019}. At such radial distances the second term on the right-hand-side of  Eq.\ref{eq:beam_intensity} is zero. It would have also been zero if the LG beam had been linearly polarized (i.e. $\sigma=0 $). These observations are telling us that at a radial distance  $\rho=w_0\sqrt{|\ell|/2}$, locally, the atom cannot distinguish whether it is interacting with a linearly or a circularly polarized beam.

It is also interesting to compare our findings with the ones presented in the work by Quinteiro et al \cite{Quinteiro2017} where they consider interaction of a tightly focused LG beam with small (dipole-like) objects that are centered  on the beam axis and so the full lateral spatial extent of the beam is irrelevant. In this case one can simplify the beam’s profile to $u \propto \rho^{|\ell|}$ (i.e. ignore the Gaussian exponential term) then Eq. \ref{eq:beam_intensity} would give $\tilde{\Omega}=\Omega\sqrt{1+\frac{1}{\rho^2}\left[\frac{1}{k\sqrt{2}}(\ell-i\sigma|\ell|)\right]^{2}}$.  The second term under the square root is non-zero only when spin and orbital angular momenta are counter-rotating. Thus our results are in perfect agreement with the result of \cite{Quinteiro2017}.

In the case of a far-off resonant interaction the relevant optical dipole potential energy is directly proportional to the intensity of the beam and thus to the square of the Rabi frequency given by:
\begin{widetext}
\bea
    U&=&\frac{\hbar\tilde{\Omega}^2}{4\Delta}\nonumber\\
    &=&\frac{\hbar}{4\Delta}\left\{\Omega^2-\frac{i\sigma \ell \Omega}{k^2\rho} \frac{\partial \Omega}{\partial \rho}+\frac{1}{2k^2}\left[\left(\frac{\partial \Omega}{\partial \rho}\right)^2+\frac{\Omega^2 \ell^2}{\rho^2}\right]\right\}
    \label{Eqn:monochromatic_potential}
\eea
\end{widetext}
Consider the case of the two-level atomic transition $6^2S_{1/2} \rightarrow 6^2P_{1/2}$ in a Cs atom with the following beam parameter values: $\ell=\pm1$, $\Delta=-7 \times 10^6\Gamma$, $\Omega_0=1.0 \times 10^4\Gamma$ (corresponding to a power $P=0.7$mW), $\lambda=852.35$nm. We assume a beam waist $w_0=0.8\lambda$;  $\Gamma=2\pi\times5.18\times10^{6} s^{-1}$ is the excited spontaneous emission rate of the upper state of the two-level atom. We then obtain for the optical dipole potential energy the  two plots shown in Fig 1. We see clearly from this figure the chiral character of the optical dipole potential energy as we reverse the sign of either the circular polarization $\sigma$ or the beam winding number $\ell$. 

\begin{figure}   
\includegraphics[width=\linewidth]{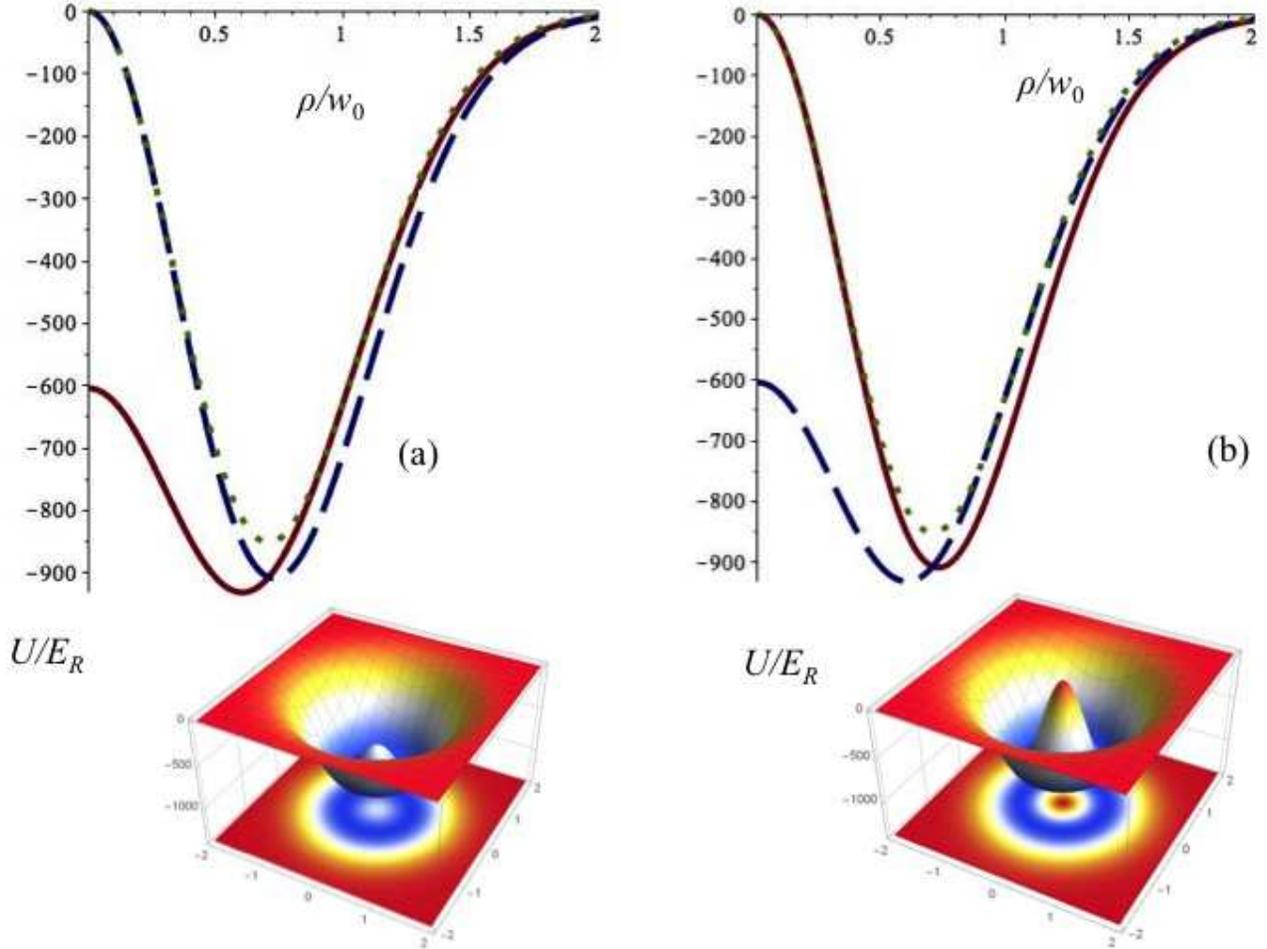}
\caption{\label{fig:Fig1.eps} (a) The optical dipole potential energy for \textbf{right-handed} circular polarization. Solid line represents the case where $\ell=+1$, dashed curve represents the case where $\ell=-1$;
(b) The optical dipole potential energy for \textbf{left-handed} circular polarization. Solid line represents the case where $\ell=-1$, dashed curve represents the case where $\ell=+1$.  In both (a) and (b) the dotted line shows the optical dipole potential energy without taking into account the contribution of the spin-orbit term. The insets to the figures show the potential energies of the solid curves and their  projections in the focal plane. These insets should be interchanged for the dashed curves. In all plots the potential energy is in recoil energy units while the radial distances are in LG beam waist $w_{0}$ units.}
\end{figure}

We also show in Fig. 2 the corresponding potential energies for the cases $\ell=\pm 5$.  In all the plots the dipole potential energy is in recoil energy units $E_R=\hbar^2k^2/2M$, where $M$ is the mass of the atom.

\begin{figure} 
\includegraphics[width=\linewidth]{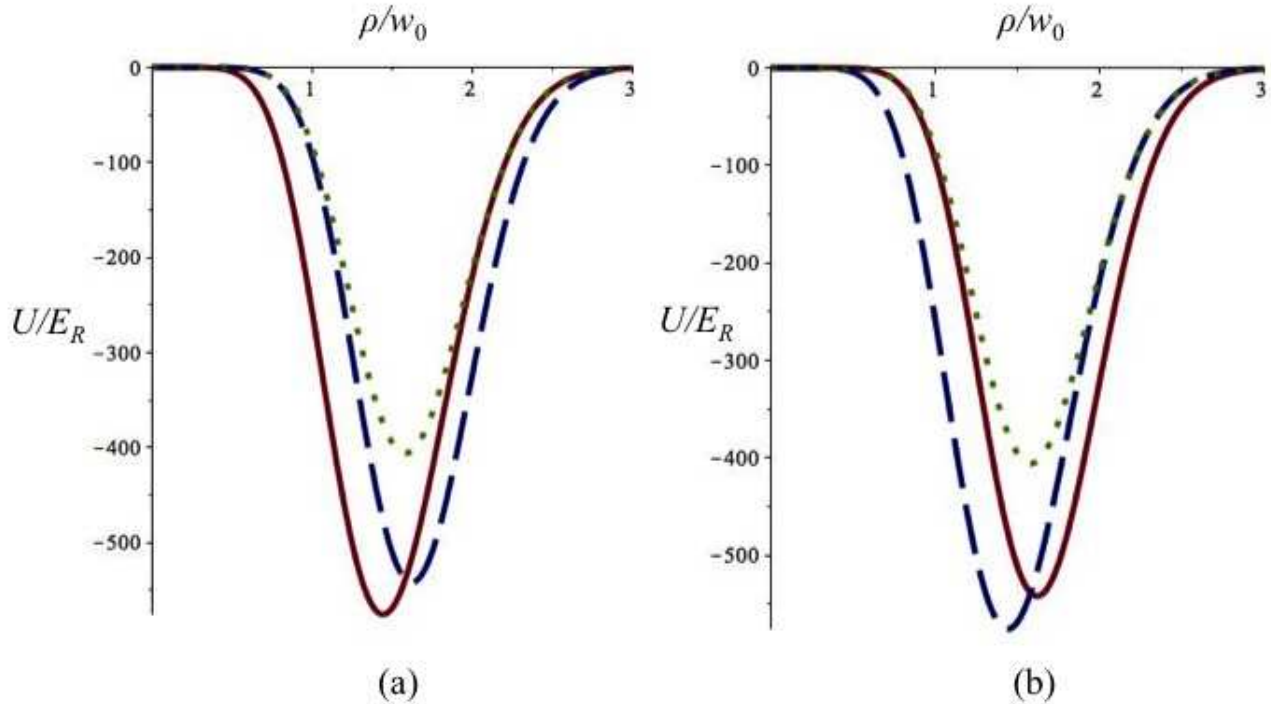}%
\caption{\label{fig:Fig2.eps} (a) The optical dipole potential energy for \textbf{right-handed} circular polarization. Solid line represents the case where $\ell=+5$, dashed curve is for $\ell=-5$. 
(b) The optical dipole potential energy for \textbf{left-handed} circular polarization. Solid line represents the case where $\ell=-5$, dashed curve is for $\ell=+5$.  In both (a) and (b) the dotted line shows the optical dipole potential energy without taking into account the contribution of the spin-orbit term. In both plots the potential energy is in recoil energy units while the radial distances are in LG beam waist $w_{0}$ units.}
\end{figure}

In Fig. \ref{fig:Fig1.eps} we see another manifestation of an effect appearing for $|\ell|=1$. When the product $\ell\sigma=1$ the atom experiences an interaction with a vortex having a vanishing intensity at the beam centre, while when $\ell\sigma=-1$ the atom experiences an interaction with a vortex having a maximum intensity at the beam centre \cite{Bliokh2015}. This effect comes directly from the helicity-dependent `switching on' of the central intensity and has been observed in experiments \cite{bokor2005, Gorodetski2008}. Note that, on carefully examining the plots, that at radial distances smaller than $\rho=w_0\sqrt{|\ell|/2}$ the chiral potential deviates from the conventional optical dipole potential for $\ell\sigma=-1$, while at radial distances larger than $\rho=w_0\sqrt{|\ell|/2}$ the chiral potential deviates from the conventional one when $\ell\sigma=1$. The explanation of this lies in Eq.(\ref{Eqn:bichromatic_potential}), that the effective Rabi frequency $\tilde{\Omega}$ becomes larger when the term $\frac{i\sigma \ell \Omega}{k^2\rho}\frac{\partial \Omega}{\partial \rho}$ is negative and this obviously depends not only on the sign of the polarization or that of the winding number but also on the gradient of the Rabi frequency.  

\begin{figure} 
\includegraphics[width=\linewidth]{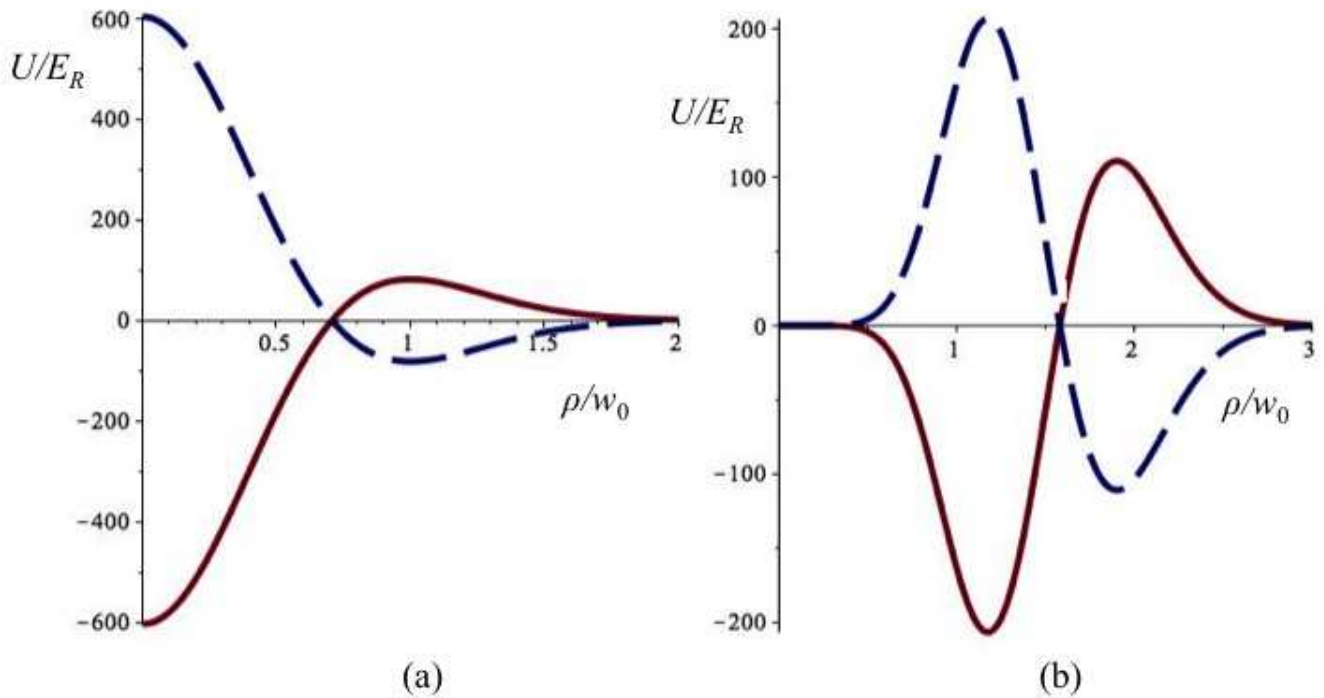}
\caption{\label{fig:Fig3.eps}  (a) The optical dipole potential energy for a bi-chromatic field with $\ell=1$ (solid line). The plot is inverted if $\ell=-1$ (dashed line). (b) Optical dipole potential energy for a bi-chromatic field with $\ell=5$ (solid line). This plot is inverted if $\ell=-5$. In both plots the potentials are given in recoil energy units while the radial distances are in LG beam waist $w_{0}$ units.}
\end{figure}

As can be seen in Eq. (\ref{Eqn:monochromatic_potential}) the optical dipole potential energy is made up of three terms the second of which is the chiral term. The question is how can we tailor the atom-beam interaction in such a way that we produce a purely chiral dipole potential? For this we consider the following bi-chromatic interaction scheme: assume two LG beams of the same winding number $\ell $, one having positive circular polarization and being blue detuned with respect to the two-level atom transition frequency $\omega_0$ (i.e $\Delta=\omega-\omega_0<0$), and the other with a negative circular polarization and being red detuned ($\Delta>0$) with respect to the two-level atom transition. Both detunings are very large in magnitude compared to the Rabi frequencies and also larger than the inverse of the excited state lifetime $\Gamma^{-1}$ so we are in the far off-resonance regime. This scenario is equivalent to the one in which the beams have the same polarization but opposite winding numbers. Note that such a scheme with bi-chromatic optical dipole potentials has been employed in the formation of a Lamb-Dicke trap for a single neutral atom \cite{He2012}, which demonstrated the existence of a state-insensitive nanofiber optical dipole trap for Cs atoms \cite{Goban2012} and the creation of an optical interface generated by cold atoms trapped around an optical nanofiber \cite{Vetsch2010}. In this case the total dipole potential for the atom is given by $U_{tot}=\frac{\hbar \tilde{\Omega}_1^2}{4\Delta}-\frac{\hbar \tilde{\Omega}_2^2}{4\Delta}$, where $\tilde{\Omega}_{1,2}^{2}=\Omega^2\mp\frac{i\sigma \ell\Omega}{k^2\rho}\frac{\partial\Omega}{\partial\rho}+\frac{1}{2k^2}\left\{\left(\frac{\partial \Omega}{\partial \rho}\right)^2+\frac{\Omega^2 \ell^2}{\rho^2}\right\}$ . Thus the total optical dipole potential experienced by the atom is:
\begin{equation}
    U=U_1+U_2=-\frac{i\hbar\sigma \ell \Omega}{2\Delta k^2\rho}\left(\frac{\partial \Omega}{\partial \rho}\right)
    \label{Eqn:bichromatic_potential}
\end{equation}
This is clearly a completely chiral optical dipole potential which obviously reverses sign if we change the sign of the winding numbers of the two beams, or if we reverse the polarization of the beams in case where we use two beams with the same polarization and opposite winding numbers. It is also obvious that it becomes zero at points where the Rabi frequency $\Omega$  (i.e. the intensity of an LG beam with a negligible longitudinal electric field component) maximizes. We present this potential in Fig \ref{fig:Fig3.eps} for two cases of different winding numbers.

\begin{figure} 
\includegraphics[width=\textwidth]{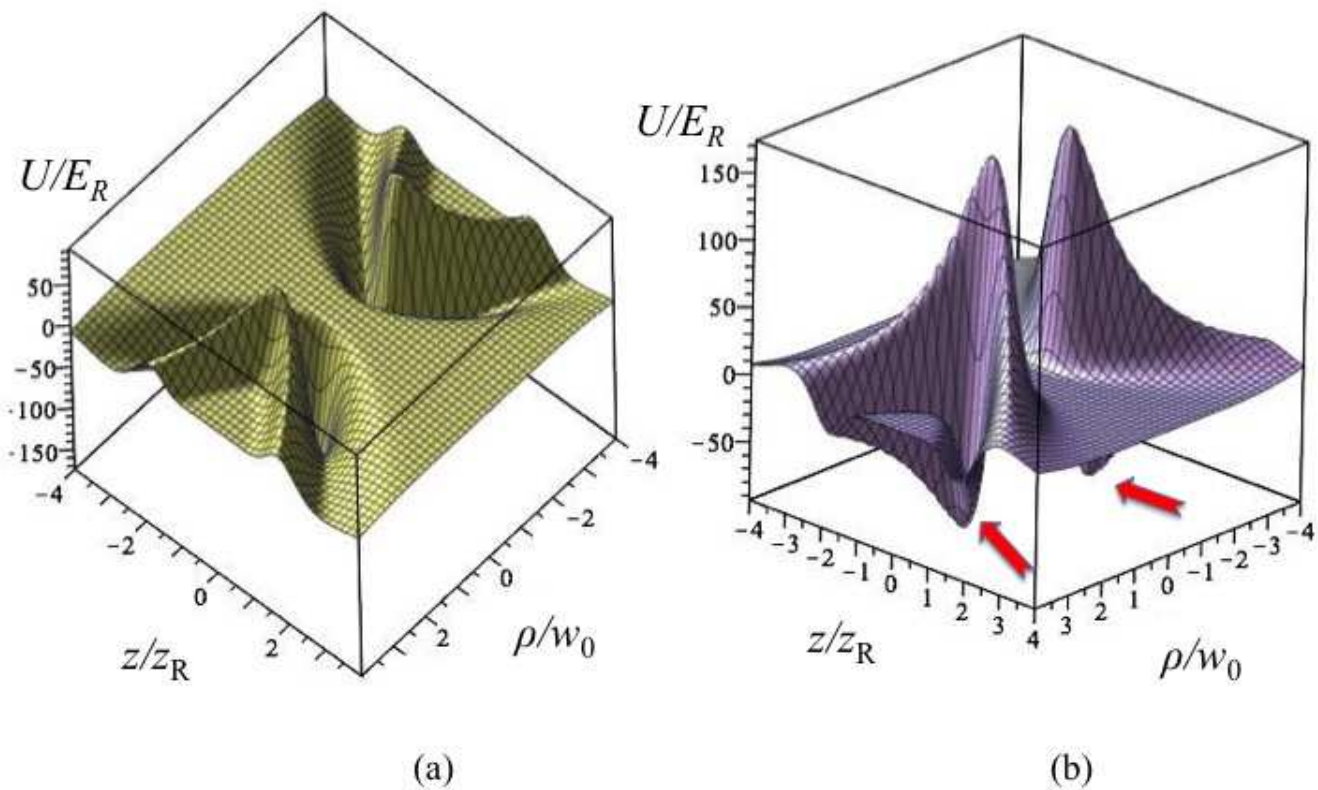}
\caption{\label{fig:Fig4.eps}  (a) The optical dipole potential energy, in the radial and axial directions, for a bi-chromatic field with $\ell=5$ (left). (b) The plot is inverted for $\ell=-5$ (right). The potential energy is given in recoil energy units, the radial distances are in LG beam waist $w_0$ units, while the axial distances are in Rayleigh range ($z_R$) units. The arrows in the plot (b) indicate the trapping regions.}
\end{figure}

In both cases presented in Fig. \ref{fig:Fig3.eps} we see clearly that the relevant potentials are quite deep and of a sub-wavelength spatial extent. It is also clear that a change in the sign of the winding number or that of the  polarization shifts the trapping position by a radial distance of the order of $w_0$. Another interesting feature is that for $\ell>1$ (Fig. \ref{fig:Fig3.eps}(b)) the dashed line case ensures two regions of trapping one of them on the axis as in a conventional Gaussian optical dipole trap while the solid line configuration gives only one trapping region.  We must emphasize that the trapping configuration which we presented do not serve only as radial trapping mechanisms but, since we assume very small beam waists, they actually operate as optical tweezers and thus can also provide axial trapping \cite{Chu1986}. For example, the three-dimensional trapping potentials for the configuration corresponding to Fig.\ref{fig:Fig3.eps}a are shown in the insets to Fig. \ref{fig:Fig4.eps}.

We see that axial trapping is possible only in the case where $\ell=5$ (left). We may have axial trapping if we exchange the signs of the detunings of the two beams keeping their polarizations fixed or if we exchange their polarizations keeping the sign of the detunings fixed.


In conclusion, we have investigated the optical dipole potential energy experienced by a two-level atom when it interacts with a circularly polarized, far off-resonant, LG beam with a small beam waist. We have found that the dipole potential, when compared to the conventional one for a broad beam illumination, includes new terms due to the existence of a longitudinal electric field component in a vortex beam with a small waist. One of the additional potential terms has a clear chiral character in the sense that it changes its sign when the sign of the circular polarization or the sign of the beam winding number is reversed and thus the atom experiences different optical dipole potentials. We also propose a bi-chromatic configuration in which the atom is illuminated by two independent beams of the same winding number with opposite circular polarizations. The beams have opposite detunings with respect to the atomic transition. In this case the resulting optical dipole potential has a purely chiral character and is sufficiently deep to trapping the atoms. Due to the small beam waist the spatial extent of the trapping regions is clearly sub-wavelength and the trapping potential depth is considerably larger than the atomic recoil energy. As we  see from Fig. 3 once we change the handedness the trapping depth is no longer symmetrical. This means that we may modify the potential depth not only by varying the detuning, as normally done in optical dipole trapping, but also  by changing the polarization adiabatically we can modify the potential depth.

\bibliography{Chiral_apssamp}

\end{document}